\documentclass[a4paper,12pt]{article}
\usepackage[dvips]{graphicx}
\usepackage[centertags]{amsmath}
\usepackage{amsfonts}
\usepackage{amssymb}
\usepackage{latexsym}
\usepackage{mathrsfs}
\usepackage{cite}
\usepackage{upgreek}

\RequirePackage{ifpdf} 
\ifpdf
  \RequirePackage[%
    unicode,%
    pdfstartview=FitH,%
    colorlinks,%
    hyperfootnotes=true,%
    bookmarksopen,%
    bookmarksnumbered,%
    bookmarksopenlevel=3%
  ]{hyperref}
  \RequirePackage{thumbpdf}
\else
  \RequirePackage{hyperref}
\fi \RequirePackage{hyperref}

\newcommand{\email}[1]{\href{mailto:#1}{\texttt{#1}}}

\newcommand{\ket}[1]{\ensuremath{\left|#1\right>}}
\newcommand{\bra}[1]{\ensuremath{\left<#1\right|}}

\newcommand{\braket}[2]{\ensuremath{\left\langle{#1}\!%
\mathrel{\left|{\vphantom{{#1} {#2}}}\right.%
\kern-\nulldelimiterspace}\!{#2}\right\rangle}}

\newcommand{\BK}[3]{\ensuremath{\left\langle{#1}\!%
\mathrel{\left|\vphantom{{#1}}{#2}\vphantom{{#3}}\right|%
\kern-\nulldelimiterspace}\!{#3}\right\rangle}}

\newcommand{\commute}[2]{\ensuremath{\left[{#1}\!%
\mathrel{\vphantom{{#1}},\vphantom{{#2}}%
\kern-\nulldelimiterspace}\!{#2}\right]}}

\newcommand{\avg}[1]{\ensuremath{\left<#1\right>}}

\newcommand{\Vacuum}{\ensuremath{\braket{0_{+}}{0_{-}}}}
\newcommand{\Vacuumm}{\ensuremath{\braket{0_{-}}{0_{-}}}}

\newcommand{\uI}{\ensuremath{\mathrm{i}}}
\newcommand{\uE}{\ensuremath{\mathrm{e}}}
\newcommand{\uD}{\ensuremath{\mathrm{d}}}

\renewcommand{\vec}[1]{\ensuremath{\boldsymbol{\mathrm{#1}}}}

\makeatletter
\let\@afterindentfalse\@afterindenttrue
\@afterindenttrue \makeatother

\begin{document}



\title{\textbf{Gravitons, induced geometry and expectation value formalism at finite temperature}}

\author{\textbf{E.~B.~Manoukian}\footnote{Corresponding author.~E-mail:~\email{manoukian\_eb@hotmail.com}.}\,
and \textbf{S.~Sukkhasena} \\
{School of Physics, \ Institute of Science,} \\
{Suranaree University of Technology,} \\
{Nakhon~Ratchasima, 30000, Thailand}}

\date{}

\maketitle

\noindent\textbf{Key~Words}~~Graviton propagator, non-conserved
external sources, quantum gravity,
expectation value formalism at finite temperature, Schwinger terms 

\begin{abstract}
After establishing the positivity constraint and spin content of
the theory for gravitons interacting with a necessarily, and
\textit{a priori}, \textit{non}-conserved external energy-momentum
tensor, the expectation value formalism of the theory is developed
at \textit{finite} temperature in the functional
\textit{differential} treatment of quantum field theory. The
necessity of having, \textit{a priori}, a non-conserved external
energy-momentum tensor is an obvious technical requirement so that
its respective ten components may be varied \textit{independently}
in order to generate expectation values and non-linearities in the
theory. The covariance of the \textit{induced} Riemann curvature
tensor, in the initial vacuum, is established even for the
quantization in a gauge corresponding only to two physical states
of the gravitons as established above. As an application, the
\textit{induced} correction to the metric and the underlying
geometry is investigated due to a closed string arising from the
Nambu action as a solution of a circularly oscillating string as,
perhaps, the simplest generalization of a limiting point-like
object. Finally it is discussed on why the geometry of spacetime
may, in general, depend on temperature due to radiative
corrections and its physical significance is emphasized.\\
\end{abstract}
\renewcommand{\theequation}{\thechapter.\arabic{equation}}
\numberwithin{equation}{section}
\section{Introduction}\label{Section1}
The graviton propagator
\cite{Manoukian_2007,Schwinger_1976,Manoukian_1990,Manoukian_1997,Manoukian_2005,Sivaram_1999}
plays a central role in the quantum field theory treatment of
gravitation. It mediates the gravitational interaction between all
particles to the leading order in the gravitational coupling
constant. It is well known that in the functional
\textit{differential} formalism of quantum field theory, pioneered
by Schwinger \cite{Schwinger_1951}, functional derivatives (e.g.,
\cite{Schwinger_1951,Manoukian_1986/1,Limboonsong_2006,Manoukian_1985,Manoukian_2006,Sukkhasena_2007})
are taken of the so-called vacuum-to-vacuum transition amplitude
$\braket{0_+}{0_-}$ with respect to external sources, via the
application, in the process, of the quantum dynamical (action)
principle (e.g.,
\cite{Manoukian_1986/1},\cite{Manoukian_2006},\cite{Sukkhasena_2007})
to generate non-linearities (interactions) in the theory and
\textit{n}-point functions leading finally to transition
amplitudes for various physical processes. [For a recent modern
and a detailed derivation of the quantum dynamical principle see
\cite{Sukkhasena_2007}.] For higher spin fields such as the
electromagnetic vector potential $A^\mu$, the gluon field
$A^\mu_a$, and, of course, the gravitational field $h^{\mu\nu}$,
the respective external sources $J_\mu$, $J^a_\mu$, $T_{\mu\nu}$,
coupled to these fields, cannot \textit{a priori} taken to be
conserved so that their respective components may be varied
\textit{independently} in the functional differentiations process.
A problem that may arise otherwise, may be readily seen from a
simple example given in \cite{Manoukian_2007}: The functional
derivative of an expression like
$\Big[a_{\mu\nu}(x)+b(x)\partial_\mu\partial_\nu\Big]T^{\mu\nu}(x)$,
with respect to $T^{\sigma\lambda}(x')$ is
$(1/2)\Big[a_{\mu\nu}(x)+b(x)\partial_\mu\partial_\nu\Big]\Big(\delta_\sigma{}^\mu
\delta_\lambda{}^\nu+\delta_\lambda{}^\mu\delta_\sigma{}^\nu\Big)\delta^4(x,x')$,
where $a_{\mu\nu}(x),b(x)$, for example, depend on \textit{x}, and
\textit{not} $(1/2)a_{\mu\nu}(x)\Big(\delta_\sigma{}^\mu
\delta_\lambda{}^\nu+\delta_\lambda{}^\mu\delta_\sigma{}^\nu\Big)\delta^4(x,x')$
as one may na\"{\i}vely assume by, \textit{a priori}, imposing a
conservation law on $T^{\mu\nu}(x)$ \textit{prior} to functional
differentiation. The consequences of relaxing the conservation of
the such external sources are highly non-trivial. For one thing
the corresponding field propagators become modified. Also they
have led to the rediscovery
\cite{Manoukian_1986/1,Limboonsong_2006} of Faddeev-Popov (FP)
\cite{Faddeev_1967}-like factors in non-abelian gauge theories
\cite{Manoukian_1986/1,Limboonsong_2006} and the discovery of even
further \textit{generalizations} \cite{Limboonsong_2006} of such
factors, directly from the functional \textit{differential}
treatment, via the application of the quantum dynamical principle
\cite{Sukkhasena_2007}, in the presence of external sources,
without making an appeal to path integrals, without using symmetry
arguments which may be broken, and without even going into the
well known complicated structures of the underlying Hamiltonians.
An account of this procedure, which is also pedagogical, was given
in the concluding section of \cite{Manoukian_2007} for the
convenience of the reader and needs not to be repeated.

For higher spin fields, the propagator and time-ordered product of
two fields do not, in general, coincide as the former includes
so-called Schwinger terms which, in general, lead to a
simplification for the propagator over the time-ordered one. This
is well known for spin 1 and is also true for the graviton
propagator \cite{Manoukian_2007}. Let $h^{\mu\nu}$ denote the
gravitational field. We work in a gauge
\begin{equation}\label{Eqn1.1}
\partial_ih^{i\nu} = 0,
\end{equation}
where $i=1,2,3;~\nu=0,1,2,3$, which, as established in Sect.3,
guarantees that only two states of polarization occur for the
graviton even with a non-conserved external source $T_{\mu\nu}$ in
the theory.

If we denote the vacuum-to-vacuum transition amplitude for the
interaction of gravitons with the external source $T_{\mu\nu}$ by
$\braket{0_+}{0_-}^T$, then the propagator of the gravitational
field is defined by
\begin{align}\label{Eqn1.2}
\Delta_+^{\mu\nu;\sigma\lambda}(x,x') = \uI\left(
(-\uI)\frac{\delta}{\delta T_{\mu\nu }(x)}
(-\uI)\frac{\delta}{\delta T_{\sigma\lambda }(x')}
\braket{0_+}{0_-}^T\right)\Big/ \braket{0_+}{0_-}^T,
\end{align}
in the limit of the vanishing of the external source $T_{\mu\nu}$.
In more detail we may rewrite (\ref{Eqn1.2}) as
\begin{align}\label{Eqn1.3}
\Delta_+^{\mu\nu;\sigma\lambda}(x,x') = \uI
\frac{\BK{0_+}{\left(h^{\mu\nu}(x)h^{\sigma\lambda}(x')\right)_+}{0_-}^T}{\braket{0_+}{0_-}^T}
+ \frac{\BK{0_+}{\dfrac{\delta}{\delta T_{\mu\nu
}(x)}h^{\sigma\lambda }(x')}{0_-}^T}{\braket{0_+}{0_-}^T}
\end{align}
in the limit of vanishing $T_{\mu\nu}$, where the first term on
the right-hand side, up to the $\uI$ factor, denotes the
time-ordered product. In the second term, the functional
derivative with respect to $T_{\mu\nu}(x)$ is taken by keeping the
independent field components of $h^{\sigma\lambda}(x')$ fixed. The
dependent field components depend on the external source and lead
to extra terms on the right-hand side of (\ref{Eqn1.3}) in
addition to the time-ordered product and may be referred to as
Schwinger terms. A detailed derivation of the general identity in
(\ref{Eqn1.3}) is given in \cite{Sukkhasena_2007} (see also
\cite{Manoukian_2006}). It is the propagator
$\Delta_+^{\mu\nu;\sigma\lambda}$ that appears in this formalism
and not the time-ordered product. The propagator
$\Delta_+^{\mu\nu;\sigma\lambda}(x,x')$ has been derived in
\cite{Manoukian_2007} and will be elaborated upon in Sect.2. It
includes 30 terms in contrast to the well known one involving only
3 terms when a conservation law of $T_{\mu\nu}$ is imposed. The
positivity constraint of the vacuum persistence probability
$|\braket{0_+}{0_-}|^2\leq1$, as well as the correct spin content
of the theory is established in Sect.3 for, \textit{a priori},
\textit{non}-conserved external energy-momentum tensor.

The expectation value formalism, pioneered by Schwinger
\cite{Schwinger_1961}, also known as the closed-time path
formalism, in quantum field theory has been a useful tool in
performing expectation values without first evaluating transition
amplitudes. For a partial list of studies of the expectation value
formalism, the reader may refer to
\cite{Manoukian_1987,Manoukian_1991/1} in the functional
differential formalism. See also related work in
\cite{Keldysh_1965,Craig_1968,Hall_1975,Kao_2002} emphasizing on
non-equilibrium phenomenae and
\cite{Jordan_1986,Calzetta_1988,Cooper_1998} emphasizing Feynman
path integrals.

In order to study gravitational effects such as the induced
geometry due to external sources and even due to fluctuating
quantum fields, the expectation value formalism turns out to be of
practical value. In Sect. 4, we develop the expectation value
formalism for gravitons interacting with an external
energy-momentum tensor $T_{\mu\nu}$ at \textit{finite} temperature
with \textit{a priori} not conserved $T_{\mu\nu}$, so that
variations with respect to its ten components may be varied
independently in order to generate expectation values.
\textit{After} all the relevant functional differentiations with
respect to $T_{\mu\nu}$ are carried out, the conservation law on
$T_{\mu\nu}$ may be then imposed. We establish the covariance of
the \textit{induced} Riemann curvature tensor, in the initial
vacuum, due to the external source, in spite of the quantization
carried out in a gauge which ensures only two polarization states
for the graviton. As an application, we investigate the
\textit{induced} correction to the metric and the underlying
geometry due a closed string arising from the Nambu action
(e.g.,\cite{Kibble_1989,Sakellariadou_1990,Goddard_1995}) as a
solution of a circularly oscillating string
\cite{Manoukian_1991/2,Manoukian_1992,Manoukian_1995,Manoukian_1998}
as, perhaps, the simplest generalization of a limiting point-like
object. Finally, it is discussed on why the geometry of spacetime
may, in general, depend on temperature due to radiative
corrections and its physical significance is emphasized. The
Minkowski metric is denoted by $[\eta_{\mu\nu}]$=diag[-1,1,1,1],
and we use units such that $\hbar=1, c=1$.

\setcounter{section}{1}

\section{Graviton propagator and vacuum-to-vacuum transition amplitude}\label{Section2}
The action for the gravitational field $h^{\mu\nu}$ coupled to an
external energy-momentum tensor source $T_{\mu\nu}$ is taken to be
\begin{align}\label{Eqn2.1}
A= \frac{1}{8\pi G}\int(\uD x)\mathscr{L}(x) + \int(\uD
x)h^{\mu\nu}(x)T_{\mu\nu}(x),
\end{align}
with
\begin{align}\label{Eqn2.2}
\mathscr{L} = -\frac{1}{2}&\partial^\alpha
h^{\mu\nu}\partial_\alpha h_{\mu\nu} + \frac{1}{2}\partial^\alpha
h^\sigma{}_\sigma \partial_\alpha h^\beta{}_\beta -
\partial^\alpha h_{\alpha\mu} \partial^\mu h^\sigma{}_\sigma \nonumber \\+ &
\frac{1}{2}\partial_\alpha h^{\alpha\nu}\partial^\beta
h_{\beta\nu} + \frac{1}{2}\partial_\alpha h^\mu{}_\nu \partial_\mu
h^{\alpha\nu} ,
\end{align}
and $G$ is Newton's gravitational constant. The action part
$\int(\uD x)\mathscr{L}(x)$ is invariant under gauge
transformations
\begin{align}\label{Eqn2.3}
h^{\mu\nu}(x) \to h^{\mu\nu}(x) + \partial^\mu\xi^\nu(x) +
\partial^\nu\xi^\mu(x) + \partial^\mu\partial^\nu\xi(x).
\end{align}
As mentioned above the external energy-momentum tensor
$T_{\mu\nu}$ is, \textit{a priori}, taken to be \textit{not}
conserved so that variations of its respective ten components may
be varied independently - a necessary \textit{technical}
requirement. Details on dependent fields due to the gauge
constraints are spelled out in \cite{Sukkhasena_2007} as well as
in \cite{Manoukian_2007}.

The vacuum-to-vacuum transition amplitude is then given by
\cite{Manoukian_2007}
\begin{align}\label{Eqn2.4}
\Vacuum^T =\exp\left[4\pi G\uI \int(\uD x)(\uD
x')T_{\mu\nu}(x)\Delta_+^{\mu\nu;\sigma\lambda}(x,x')T_{\sigma\lambda}(x')\right],
\end{align}
\begin{align}\label{Eqn2.5}
(\uD x) = \uD x^0\uD x^1 \uD x^2 \uD x^3.
\end{align}
Here we note that the exponent is scaled by the factor $8\pi G$ to
satisfy the boundary condition that the gravitational attraction
of two widely separated static sources is given by Newton's law
\cite{Schwinger_1976}. The graviton propagator
$\Delta_+^{\mu\nu;\sigma\lambda}(x,x')$ contains 30 terms and
\textit{not} only just the first 3 terms as may be na\"{\i}vely
expected, and is given by
\begin{align}\label{Eqn2.6}
\Delta_+^{\mu\nu;\sigma\lambda}(x,x')  = \int\dfrac{(\uD
k)}{(2\pi)^4}\uE^{\uI
k(x-x')}\left[\frac{\Delta_1^{\mu\nu;\sigma\lambda}(k)}{k^2-\uI\epsilon}
+ \frac{\Delta_2^{\mu\nu;\sigma\lambda}(k)}{\vec{k}^2}\right],
\end{align}
$\epsilon\rightarrow+0$, where $(\uD k)=\uD k^0 \uD k^1 \uD k^2
\uD k^3$, $k^2=\vec{k}^2-k^{0^2}$, and
\begin{align}\label{Eqn2.7}
\Delta_1^{\mu\nu;\lambda\sigma}(k) =
\frac{(\eta^{\mu\lambda}\eta^{\nu\sigma} +
\eta^{\mu\sigma}\eta^{\nu\lambda} -
\eta^{\mu\nu}\eta^{\sigma\lambda})}{2}&\nonumber \\
        +\frac{1}{2\vec{k}^2}\Big[\eta^{\mu\nu}k^\sigma k^\lambda +
        \eta^{\sigma\lambda}k^\mu k^\nu &-\eta^{\nu\sigma}k^\mu k^\lambda
        -\eta^{\nu\lambda}k^\mu k^\sigma  \nonumber\\
-\eta^{\mu\sigma}k^\nu k^\lambda -\eta^{\mu\lambda}k^\nu k^\sigma
&+ \frac{k^{\mu}k^{\nu}k^{\sigma}k^{\lambda}}{\vec{k}^2}\Big]
\nonumber\\
        -\frac{1}{2}\left(\eta^{\mu\nu} +
        \frac{k^{\mu}k^\nu}{\vec{k}^2}\right)\left(\frac{N^\sigma{}k^\lambda
        + N^\lambda{}k^\sigma}{\vec{k}^2}\right)k^0 & \nonumber\\
-\frac{1}{2}\left(\eta^{\sigma\lambda} +
\frac{k^\sigma{}k^\lambda}{\vec{k}^2}\right)\left(\frac{N^\nu{}k^\mu
+N^\mu{}k^\nu}{\vec{k}^2}\right)k^0&\nonumber \\
    +\frac{1}{2}\Big[\eta^{\nu\sigma}(N^\mu{}k^\lambda + N^\lambda{}k^\mu)
    + \eta^{\nu\lambda}(N^\mu{}k^\sigma + N^\sigma{}k^\mu)&\nonumber \\
        +\eta^{\mu\sigma}(N^\nu{}k^\lambda + N^\lambda{}k^\nu) +
        \eta^{\mu\lambda}(N^\nu{}k^\sigma &+
        N^\sigma{}k^\nu)\Big]\frac{k^0}{\vec{k}^2}\nonumber\\
+\frac{k^\mu k^\nu}{\vec{k}^2}N^\sigma N^\lambda + \frac{k^\sigma
k^\lambda}{\vec{k}^2} N^\mu N^\nu &,
\end{align}
\begin{align}\label{Eqn2.8}
\Delta_2^{\mu\nu;\lambda\sigma}(k)  =  \frac{k^\mu
k^\nu}{\vec{k}^2} N^\sigma N^\lambda + \frac{k^\sigma
k^\lambda}{\vec{k}^2} N^\mu N^\nu.
\end{align}
Here $(N^\mu)=(\eta^\mu{}_0)=(1,0,0,0)$. The $\uI\epsilon$ factor
in (\ref{Eqn2.6}) corresponds to the Schwinger-Feynman boundary
condition.

It is far from obvious that with a \textit{non-conserved}
energy-momentum tensor, the vacuum-to-vacuum amplitude
$\braket{0_+}{0_-}$ in (\ref{Eqn2.4}) satisfies the positivity
constraint $|\braket{0_+}{0_-}|^2\leq1$. This together with the
correct spin content of the theory is established in the next
section.

\section{Positivity constraint and spin content}\label{Section3}

We rewrite the vacuum-to-vacuum transition amplitude
$\braket{0_+}{0_-}$ in (\ref{Eqn2.4}) as
\begin{align}\label{Eqn3.1}
\Vacuum^T = \exp\left[4\pi G\uI \int (\uD
x)T_{\mu\nu}(x)H^{\mu\nu}(x)\right],
\end{align}
with
\begin{align}\label{Eqn3.2}
T_{\mu\nu}H^{\mu\nu} = T_{00}H^{00} + 2T_{0i}H^{0i} +
T_{ij}H^{ij},
\end{align}
$i,j=1,2,3$, and we may infer from Eq.(13) in
\cite{Manoukian_2007} that
\begin{align}\label{Eqn3.3}
H^{00} = -\frac{1}{\vec{\partial}^2}\left[T^{00}
        + \frac{T}{2}
        -\frac{1}{2\vec{\partial}^2}\left(\partial^0\partial^0 T_{00}
        +\partial^i\partial^j T_{ij}\right)\right],
\end{align}
$T=T_{ii}-T_{00}$, and $H^{00}$ is \textit{real}. Also from
Eq.(12) in \cite{Manoukian_2007}, we may infer that
\begin{align}\label{Eqn3.4}
H^{0i} = -\frac{1}{\vec{\partial}^2}\left[\updelta^{ij} -
\frac{\partial^i\partial^j}{\vec{\partial}^2}\right]T_{0j},
\end{align}
which is again \textit{real}. That is,
\begin{align}\label{Eqn3.5}
\exp\left[4\pi G\uI \int (\uD x)\left(T_{00}(x)H^{00}(x) +
2T_{0i}(x)H^{0i}(x)\right)\right]
\end{align}
is a phase factor.

On the other hand, we may infer from Eq.(17) in
\cite{Manoukian_2007} that
\begin{align}\label{Eqn3.6}
H^{ij} = \frac{1}{(-\Box -\uI\epsilon)}A^{ij,lm}T_{lm}
        - \frac{1}{2}\frac{1}{\vec{\partial}^2}\left(\updelta^{ij}
        - \frac{\partial^i\partial^j}{\vec{\partial}^2}\right)T_{00},
\end{align}
and the second term above involving $T_{00}$ is \textit{real},
while $A^{ij,lm}$ is given by
\begin{align}\label{Eqn3.7}
A^{ij,lm} &= \frac{(\updelta^{il}\updelta^{jm}
            +\updelta^{im}\updelta^{jl}
            -\updelta^{ij}\updelta^{lm})}{2}\nonumber\\
&-\frac{1}{2\vec{\partial}^2}\bigg[
\partial^i\partial^l\updelta^{jm}
+\partial^i\partial^m\updelta^{jl} +
\partial^j\partial^l\updelta^{im} +
\partial^j\partial^m\updelta^{il}\nonumber\\ &+
\partial^i\partial^j\updelta^{lm}-
\updelta^{ij}\partial^l\partial^m -
\frac{\partial^i\partial^j\partial^l\partial^m}{\vec{\partial}^2}
\bigg]
\end{align}
where $i,j,l,m=1,2,3$.

Accordingly, from (\ref{Eqn3.1}), (\ref{Eqn3.5})-(\ref{Eqn3.7}),
we may rewrite
\begin{align}\label{Eqn3.8}
\Vacuum^T =\uE^{\uI G[T]}\exp\left[4\pi G\uI \int(\uD
x)T_{ij}(x)\frac{1}{(-\Box -
\uI\epsilon)}A^{ij,lm}T_{lm}(x)\right],
\end{align}
where $\exp\uI G[T]$ is a phase factor.

By using the facts that the reality of $T_{ij}(x)$ implies that
$T_{ij}(k)^*=T_{ij}(-k)$, where $(k^\mu)=(k^0,\vec{k})$, and the
identity
\begin{align}\label{Eqn3.9}
\frac{\uI}{2}\left(\frac{1}{k^2-\uI\epsilon}
                            - \frac{1}{k^2
                    +\uI\epsilon}\right)
                        = -\pi\delta(k^2)
                        = -\frac{\pi}{|\vec{k}|}
                            \Big[\delta(k^0 - |\vec{k}|) + \delta(k^0 + |\vec{k}|)\Big]
\end{align}
for $\epsilon\rightarrow+0$, in the sense of distributions, we
obtain that
\begin{align}\label{Eqn3.10}
\left|\Vacuum^T\right|^2 = \exp\left[-8\pi G \int\uD
\omega_{\vec{k}}T^*_{ij}(k)B^{ij,lm}(k)T_{lm}(k)\right],
\end{align}
where now $k^0=+|\vec{k}|$,
$\uD\omega_{\vec{k}}=\uD^3\vec{k}/(2\pi)^32|\vec{k}|$, and
\begin{align}\label{Eqn3.11}
B^{ij,lm}(k) = &\frac{1}{2}
                \bigg[
                (\updelta^{il}-\frac{k^i k^l}{\vec{k}^2})
                (\updelta^{jm}-\frac{k^j k^m}{\vec{k}^2})
                +(\updelta^{im}-\frac{k^i k^m}{\vec{k}^2})
                (\updelta^{jl}-\frac{k^j
                k^l}{\vec{k}^2})\nonumber\\
                &-(\updelta^{ij}-\frac{k^i k^j}{\vec{k}^2})
                (\updelta^{lm}-\frac{k^l k^m}{\vec{k}^2})
                \bigg],
\end{align}
with $i,j,l,m=1,2,3$ as before.

For a given 3-vector $\vec{k}$, we introduce two orthonormal
complex 3-vectors $\vec{e}_+,\vec{e}_-$,
\begin{align}\label{Eqn3.12}
\vec{e}_+\cdot\vec{e}^*_+ = 1 = \vec{e}_-\cdot\vec{e}^*_- ,\qquad
\vec{e}_+\cdot\vec{e}^*_- = 0,
\end{align}
such that $\vec{k}/|\vec{k}|$, $\vec{e}_+,\vec{e}_-$ constitute
three mutually orthonormal vectors. That is, in addition to the
conditions in (\ref{Eqn3.12}),
\begin{align}\label{Eqn3.13}
\vec{k}\cdot \vec{e}_+ = 0,\qquad \vec{k}\cdot\vec{e}_-= 0.
\end{align}
Upon writing
\begin{align}\label{Eqn3.14}
\vec{k} = |\vec{k}|
            \Big(\cos\phi\sin\theta, \sin\phi\sin\theta,
            \cos\theta\Big),
\end{align}
we may set
\begin{align}\label{Eqn3.15}
\vec{e}_+ = \frac{1}{\sqrt{2}}
            \Big(\cos\phi\cos\theta -\uI\sin\phi,
                \sin\phi\cos\theta
                + \uI\cos\phi,
                -\sin\theta\Big),
\end{align}
\begin{align}\label{Eqn3.16}
\vec{e}_- = \frac{1}{\sqrt{2}}
            \Big(\cos\phi\cos\theta +\uI\sin\phi,
            \sin\phi\cos\theta - \uI\cos\phi,
            -\sin\theta\Big),
\end{align}

and note that
\begin{align}\label{Eqn3.17}
\vec{e}_- = \vec{e}^*_+\quad.
\end{align}

The above allows us to introduce the completeness relation
\begin{align}\label{Eqn3.18}
\updelta^{ij} &= \sum_{\lambda=\pm} e^i_\lambda e^{*j}_\lambda +
\frac{k^i k^j}{|\vec{k}|^2} \nonumber\\
&=\sum_{\lambda=\pm} e^{i*}_\lambda e^{j}_\lambda + \frac{k^i
k^j}{|\vec{k}|^2}
\end{align}
In turn, we may define polarization 3x3 tensors by
\begin{align}\label{Eqn3.19}
e^{ij}_{\lambda\sigma} = \frac{1}{2}
                        \Big[e^i_\lambda e^{j*}_\sigma{}+e^{i*}_\sigma e^j_{\lambda}
                        -\updelta_{\lambda\sigma}e^{i}_{\alpha}e^{j*}_\alpha\Big]
\end{align}
with $\lambda,\sigma,\alpha=\pm$, and a summation over the
repeated index $\alpha$ is assumed, and note that after some
algebra, $B^{ij,lm}$ in (\ref{Eqn3.11}) may be rewritten as
\begin{align}\label{Eqn3.20}
B^{ij,lm} = \sum_{\lambda,\sigma = \pm}
e^{ij}_{\lambda\sigma}e^{*lm}_{\lambda\sigma}
\end{align}

Using, in the process, (\ref{Eqn3.19}), we note that
\begin{align}\label{Eqn3.21}
e^{i\,j}_{++} = 0 , \qquad  e^{i\,j}_{--} = 0,
\end{align}
and
\begin{align}
e^{i\,j}_{+-} = e^{i}_+ e^{j}_+ \equiv \ \epsilon^{ij}_+,\label{Eqn3.22}\\
e^{i\,j}_{-+} = e^{i}_- e^{j}_- \equiv \
\epsilon^{ij}_-\label{Eqn3.23}
\end{align}
thus defining the  two 3x3 tensors $\ \epsilon_+^{ij},\
\epsilon_-^{ij}$, and rewrite (\ref{Eqn3.20}) as
\begin{align}\label{Eqn3.24}
B^{ij,lm} = \sum_{\lambda = \pm}\,\, \ \epsilon^{ij}_\lambda \
\epsilon^{lm^*}_\lambda
\end{align}

From (\ref{Eqn3.10}), (\ref{Eqn3.11}), (\ref{Eqn3.24}), we
conclude that
\begin{align}\label{Eqn3.25}
\left|\Vacuum^T\right|^2 = \exp\Big[-8\pi G \int
\uD\omega_{\vec{k}}
\sum_{\lambda=\pm}\left(T^*_{ij}\epsilon^{ij}_\lambda\right)\left(\epsilon^{lm*}_{\lambda}T_{lm}\right)\Big]
 \leq 1
\end{align}
with equality holding in the limit of vanishing $T_{\mu\nu}$, thus
establishing the underlying positivity constraint, as well as the
correct spin content of the theory with the graviton having only
two polarization states described by
$\epsilon_+^{ij},\epsilon_-^{ij}$ for a theory with, in general, a
\textit{not} necessarily conserved external energy-momentum
tensor.

The scalar product in (\ref{Eqn3.25}) may be rewritten from
(\ref{Eqn3.24}) as follows
\begin{align}\label{Eqn3.26}
\int \uD\omega_{\vec{k}}\sum_{\lambda=\pm}
(T^*_{ij}\epsilon^{ij}_\lambda)(\epsilon^{lm*}_\lambda T_{lm}) &=
\int \uD\omega_{\vec{k}} T^*_{ij}B^{ij,lm}T_{lm}
\nonumber\\
&=\int (\uD x)(\uD x') T_{\mu\nu}(x)
C^{\mu\nu,\sigma\rho}(x,x')T_{\sigma\rho}(x'),
\end{align}
where
\begin{align}
C^{\mu\nu,\sigma\rho}(x,x')&=\int\uD\omega_{\vec{k}}\uE^{\uI k(x-x')}\pi^{\mu\nu,\sigma\rho}(k),\label{Eqn3.27}\\[0.5\baselineskip]
\pi^{\mu\nu,\sigma\rho}(k)&=\frac{1}{2}\left(\beta^{\mu\sigma}\beta^{\nu\rho}
                            + \beta^{\mu\rho}\beta^{\nu\sigma}
                            -\beta^{\mu\nu}\beta^{\sigma\rho}
                            \right),\label{Eqn3.28}\\[0.5\baselineskip]
\beta^{\mu\nu}(k)&=\left[\eta^{\mu\nu}-
\frac{k^{\mu}k^{\nu}}{(Nk)^2}
                   -\frac{N^\mu k^\nu}{(Nk)}
                    -\frac{N^\nu k^\mu}{(Nk)}\right],\label{Eqn3.29}\\[0.5\baselineskip]
Nk&= N_{\alpha}k^{\alpha} = -k^0 = - |\vec{k}|.\label{Eqn3.30}
\end{align}

\section{Gravitons and expectation value formalism at finite temperature}\label{Section4}
For book-keeping purposes, we use the notation
\begin{align}\label{Eqn4.1}
\sqrt{8\pi G}\,\epsilon^{lm*}_{\lambda} T_{lm}(\vec{k}) \equiv
S(\vec{k},\lambda),
\end{align}
and conveniently introduce a discrete notation
\cite{Schwinger_1976,Manoukian_1986/2} for the momentum variable
$\vec{k}$ by writing, in the process, $(\vec{k},\lambda)\equiv r$
for these pairs of variables and in turn use the notation $S_r$
for $S(\vec{k},\lambda)$. A scalar product as in (\ref{Eqn3.25})
then becomes simply replaced as follows:
\begin{align}\label{Eqn4.2}
8\pi G \int\uD\omega_{\vec{k}}
\sum_{\lambda=\pm}\left(T^*_{ij}\epsilon^{ij}_{\lambda}\right)\left(\epsilon^{lm*}_{\lambda}T_{lm}
\right) \to \sum_r S^*_r S_r.
\end{align}

With the above notation, and for any two, \textit{a priori},
independent, not necessarily conserved, sources $T^1_{\mu\nu}$,
$T^2_{\mu\nu}$, we introduce the functional
\begin{align}\label{Eqn4.3}
\mathscr{F}[T^1,T^2] = \sum_N \sum_{N_1 + N_2 + ... = N}
\braket{0_-}{N;N_1,N_2,...}^{T^2}\braket{N;N_1,N_2,...}{0_-}^{T^1},
\end{align}
where $N$ denotes number of gravitons, $N_1$ of which have
momentum-polarization index $r_1$, and so on, with
$\braket{N;N_1,N_2,\ldots}{0_-}^{T^1}$ denoting the amplitude that
these $N$ gravitons are emitted by the source $T^1$, and is given
by
\begin{align}\label{Eqn4.4}
\braket{N;N_1,N_2,...}{0_-}^{T^1} = \braket{0_+}{0_-}^{T^1}
\frac{(\uI S^1_{r_1})^{N_1}} {\sqrt {N_1}!} \frac{(\uI
S^1_{r_2})^{N_2}}{\sqrt {N_2}!}...~.
\end{align}

The expression for the functional $\mathscr{F}[T^1,T^2]$ may be
summed exactly by using, in the precess, (\ref{Eqn4.4}), to give
\begin{align}\label{Eqn4.5}
\mathscr{F}[T^1,T^2] &=
                        \left(\Vacuum^{T^2}\right)^*\left(\Vacuum^{T^1}\right)\nonumber\\
            &\quad\times\exp\bigg[8\pi G\int\uD
            \omega_{\vec{k}}\sum_{\lambda=\pm}
            \left(T^{*2}_{ij}\epsilon^{ij}_{\lambda}\right)
            \left(\epsilon^{lm*}_{\lambda}T^1_{lm}\right)\bigg],
\end{align}
where we have restored the integration signs. From (\ref{Eqn4.3}),
we realize that for the special case that $T^1_{\mu\nu}$ and
$T^2_{\mu\nu}$ are equal, we have by unitarity
\begin{align}\label{Eqn4.6}
\mathscr{F}[T,T] = \Vacuumm^T = 1
\end{align}
which also follows readily from (\ref{Eqn4.5}) and the left-hand
side equality in (\ref{Eqn3.25}).

In the expression for $\mathscr{F}[T^1,T^2]$, we write
$T^1=T_1+T'_1$, $T^2=T_2+T'_2$, where $T'_1$ is switched on after
$T_1$ is switched off, and $T'_2$ is switched on after $T_2$ is
switched off, to obtain from (\ref{Eqn4.3}) and (\ref{Eqn4.5}),
\textit{respectively},
\begin{align}\label{Eqn4.7}
&\mathscr{F}[T_1 + T'_1, T_2 + T'_2]
=\sum_{(N)}\braket{0_-}{N;N_1,N_2,\ldots}^{T_2 + T'_2}
\braket{N;N_1,N_2,\ldots}{0_-}^{T_1+ T'_1}\nonumber\\
&=\sum_{(N),(M)}\braket{0_-}{N;N_1,N_2,\ldots}^{T_2}\braket{N;N_1,N_2,\ldots}{M;M_1,M_2,\ldots}^{T'_2,T'_1}\nonumber\\
&\quad\times\braket{M;M_1,M_2,\ldots}{0_-}^{T_1},
\end{align}
where
\begin{align}\label{Eqn4.8}
\braket{N;N_1,N_2,\ldots}{M;M_1,M_2,\ldots}^{T'_2,T'_1}
=\sum_{(L)}\braket{N;N_1,N_2,\ldots}{L;L_1,L_2,\ldots}^{T'_2}\nonumber\\
\times\braket{L;L_1,L_2,\ldots}{M;M_1,M_2,\ldots}^{T'_1},
\end{align}
with $\underset{(N)}{\sum}$ denoting a sum over non-negative
integers $N, N_1, N_2,\ldots$ such that $N_1+N_2+\ldots=N$, and
similarly for $\underset{(M)}{\sum}$\,, $\underset{(L)}{\sum}$\,,
\textit{and}
\begin{align}\label{Eqn4.9}
\mathscr{F}[T_1 + T'_1, T_2 + T'_2] = \mathscr{F}[T'_1
,T'_2]\exp[S^*_2
S_1]\left(\Vacuum^{T_2}\right)^*\left(\Vacuum^{T_1}\right)\nonumber\\
\times\exp[S^*_2 (S'_1 - S'_2)]\exp[-(S^{'*}_1 - S^{'*}_2)S_1],
\end{align}

\noindent where the scalar product $S^*_2S_1$, for example, is
defined as on the right-hand side of (\ref{Eqn4.2}) with a sum
over $r$. Upon comparison of the two equivalent expressions for
$\mathscr{F}[T_1+T'_1,T_2+T'_2]$ in (\ref{Eqn4.7}) and
(\ref{Eqn4.9}), we obtain, in particular, for the diagonal term
$\braket{N; N_1, N_2,\ldots}{N; N_1, N_2,\ldots}^{T^2,T^1}$, valid
for \textit{any} two, \textit{a priori}, independent and not
necessarily conserved sources $T^1_{\mu\nu}, T^2_{\mu\nu}$, the
expression:
\begin{align}\label{Eqn4.10}
\braket{N;N_1,N_2,\ldots}{N;N_1,N_2,\ldots}^{T^2,T^1} =
(N_1!N_2!\cdots)\mathscr{F}[T^1,T^2]\nonumber\\
\times \sum{}^*\prod_i
    \frac{\left[-(S^{1*}_{r_i}-S^{2*}_{r_i})(S^1_{r_i}-S^2_{r_i})\right]^{N_i-m_i}}
         {m_i![(N_i - m_i)!]^2},
\end{align}
where $\sum^*$ stands for a summation over all non-negative
integers $m_1, m_2,\ldots$ such that $0\leq m_i\leq N_i$,
$i=1,2,\ldots$.

We now perform a thermal average \cite{Manoukian_1991/1} of
$\braket{N; N_1, N_2,\ldots}{N; N_1, N_2,\ldots}^{T^2,T^1}$ by
multiplying, in the process, the latter by the Boltzmann factor
$\underset{i}{\prod}(\exp-\beta|\vec{k}_i|)$ and summing over
$(N)$, where $\beta=1/\textrm{K}\tau$, and we have used the
notation $\textrm{K}$ for the Boltzmann constant and $\tau$ for
temperature in order not to confuse it with the trace $T$ of an
energy-momentum tensor. This gives the  statistical thermal
average:
\begin{align}\label{Eqn4.11}
&\avg{\mathscr{F}[T^1,T^2]}_\tau \equiv
\mathscr{F}[T^1,T^2;\tau]\nonumber\\
&=\mathscr{F}[T^1,T^2;0]
    \exp\Big[-8\pi G \int\uD
            \omega_{\vec{k}}\sum_{\lambda=\pm}
            \frac{(T^{1*}_{ij} - T^{2*}_{ij})\epsilon^{ij}_{\lambda}\epsilon^{lm*}_{\lambda}(T^1_{lm}-T^2_{lm})}
            {(\uE^{\beta |\vec{k}|} - 1)}\Big]
\end{align}

In particular, we note from (\ref{Eqn4.5}), (\ref{Eqn4.6}),
(\ref{Eqn4.11}) that for the special case that $T^1_{\mu\nu}$,
$T^2_{\mu\nu}$ are identical, we have the consistent normalization
condition
\begin{align}\label{Eqn4.12}
\mathscr{F}[T,T;\tau] \equiv 1.
\end{align}

We also verify directly from (\ref{Eqn4.11}) that
\begin{align}\label{Eqn4.13}
\mathscr{F}[T^1,T^2;0] = \mathscr{F}[T^1,T^2],
\end{align}
as expected.

As we have not imposed conservation laws on $T^1_{\mu\nu}$,
$T^2_{\mu\nu}$, we may vary each of their respective ten
components independently to obtain from the quantum dynamical
principle \cite{Schwinger_1951,Sukkhasena_2007,Schwinger_1961} as
applied, respectively, and in the process to $\braket{L;
L_1,\ldots}{M; M_1,\ldots}^{T^1}$ and $\braket{N; N_1,\ldots}{L;
L_1,\ldots}^{T^2}$ in (\ref{Eqn4.8}) with $T'_1, T'_2$ in it
replaced by $T^1, T^2$, the thermal average
$\avg{h^{\mu\nu}(x)}^T_\tau$ of the gravitational field
\begin{align}\label{Eqn4.14}
\avg{h^{\mu\nu}(x)}^T_\tau &= (-\uI)\frac{\updelta}{\updelta
                                T^1_{\mu\nu}(x)}\mathscr{F}[T^1,T^2;\tau]\bigg|_{T^1=T^2=T}\nonumber\\
                           &=(\uI)\frac{\updelta}{\updelta
                                T^2_{\mu\nu}(x)}\mathscr{F}[T^1,T^2;\tau]\bigg|_{T^1=T^2=T}\quad,
\end{align}
generalizing the expression for
$\bra{0_-}h^{\mu\nu}(x)\ket{0_-}^T$ given by
\begin{align}\label{Eqn4.15}
\BK{0_-}{h^{\mu\nu}(x)}{0_-}^T &= (-\uI)\frac{\updelta}{\updelta
                                T^1_{\mu\nu}(x)}\mathscr{F}[T^1,T^2]\bigg|_{T^1=T^2=T}\nonumber\\
                           &=(\uI)\frac{\updelta}{\updelta
                                T^2_{\mu\nu}(x)}\mathscr{F}[T^1,T^2]\bigg|_{T^1=T^2=T}\quad,
\end{align}
from zero to finite temperature.

From (\ref{Eqn4.11}), (\ref{Eqn4.5}), (\ref{Eqn3.26}), the
generating functional $\mathscr{F}[T^1,T^2;\tau]$ may be rewritten
as
\begin{align}\label{Eqn4.16}
\mathscr{F}[T^1,T^2;\tau] &=
                            (\Vacuum^{T^2})^*(\Vacuum^{T^1})\nonumber\\
                       &\times \exp\left[8\pi G \int (\uD x)(\uD
                       x')T^2_{\mu\nu}(x)C^{\mu\nu,\sigma\rho}(x,x')T'_{\sigma\rho}(x')\right]\nonumber\\
    &\times\exp
    \bigg[-8\pi G \int (\uD x)(\uD x')
        (T^1_{\mu\nu}(x)-T^2_{\mu\nu}(x'))\nonumber\\
        &\qquad\times D^{\mu\nu,\sigma\rho}(x,x';\tau)
        (T^1_{\sigma\rho}(x') - T^2_{\sigma\rho}(x'))
    \bigg],
\end{align}
where $C^{\mu\nu,\sigma\rho}(x,x')$ is defined in (\ref{Eqn3.27}),
and
\begin{align}\label{Eqn4.17}
D^{\mu\nu,\sigma\rho}(x,x';\tau) = \int\uD\omega_{\vec{k}}
                                    \uE^{\uI k(x-x')}
                                    \frac{\pi^{\mu\nu,\sigma\rho}(k)}{(\uE^{-\beta(Nk)}-1)},
\end{align}
$Nk=N_\alpha k^\alpha=-k^0=-|\vec{k}|$, where
$\pi^{\mu\nu,\sigma\rho}(k)$ is given in (\ref{Eqn3.28}).

We note that the temperature dependence occurs only in the last
exponential in (\ref{Eqn4.16}) through
$D^{\mu\nu,\sigma\rho}(x,x';\tau)$. We eventually set
$T^1_{\mu\nu}=T^2_{\mu\nu}$ \textit{after} the relevant functional
differentiations with respect to these sources are taken. For
$\tau\rightarrow 0$, the last exponential in (\ref{Eqn4.16}) is
equal to one, giving the relation in (\ref{Eqn4.13}).

\section{Covariance of the induced Riemann curvature tensor}\label{Section5}

The thermal average $\avg{h_{\mu\nu}(x)}^T_\tau$ may be obtained
from (\ref{Eqn4.14}), (\ref{Eqn4.16}) to give
\begin{align}\label{Eqn5.1}
\avg{h_{\mu\nu}(x)}^T_\tau &= 8\pi G\uI \int (\uD x')
                                T^{\sigma\rho}(x') \int \uD
                                \omega_{\vec{k}}\pi_{\mu\nu,\sigma\rho}(k)
                                \uE^{\uI k(x-x')}\nonumber\\
                           &\qquad-8\pi G\uI \int (\uD
                           x')T^{\sigma\rho}(x') \int \uD\omega_{\vec{k}}\pi_{\sigma\rho,\mu\nu}(k)
                                \uE^{\uI k(x'-x)}\nonumber\\
        &=-16\pi G \int(\uD x')T^{\sigma\rho}(x')
        \int \uD\omega_{\vec{k}}\sin
        k(x-x')\pi_{\mu\nu,\sigma\rho}(k)\nonumber\\
        &\equiv \BK{0_-}{h_{\mu\nu}(x)}{0_-}^T
\end{align}
for $x^0>x'^0$, where after the functional differentiation was
carried out with respect to, say, $T^{1\mu\nu}(x)$, we have set
$T^{2\mu\nu}=T^{1\mu\nu}=T^{\mu\nu}$. We learn that the above
expectation value is independent of temperature in the leading
linearized theory as a consequence of the fact that the exponent
in the last exponential in (\ref{Eqn4.16}) does not contribute if
a single functional differentiation w.r.t. $T^{1\mu\nu}$ is
carried out and then by finally setting
$T^2_{\mu\nu}-T^1_{\mu\nu}=0$. Radiative corrections and explicit
temperature dependence will be discussed in Sect.\ref{Section7}.

In more detail, we may rewrite (\ref{Eqn5.1}) as:
\begin{align}\label{Eqn5.2}
\BK{0_-}{h_{\mu\nu}(x)}{0_-}^T &= \left\{8\pi G\uI \int
                                    \uD\omega_{k} \uE^{\uI kx}
                                    \left[T_{\mu\nu}(k)-\frac{g_{\mu\nu}}{2}T(k)\right]
                                    + c.c.\right\}\nonumber\\
                                &\quad+ \partial_{\mu}\xi_{\nu}(x)
                                 + \partial_{\nu}\xi_{\mu}(x)
                                 + \partial_{\mu}\partial_{\nu}\xi(x)
\end{align}
for $x^0 > x'^0$, where
\begin{align}
 \xi_{\mu}(x) = \left\{4\pi G
                \int \uD\omega_{\vec{k}} \uE^{\uI kx}
                    \frac{N_{\mu}T - 2T_{\mu}{}^{\sigma}N_{\sigma}}
                         {(Nk)} + c.c.\right\}\label{Eqn5.3}\\[0.5\baselineskip]
\xi(x) = \left\{\frac{4\pi G}{\uI}
                \int \uD\omega_{\vec{k}} \uE^{\uI kx}
                    \frac{T + 2T^{\nu\sigma}N_{\nu}N_{\sigma}}
                         {(Nk)^2} + c.c.\right\}\label{Eqn5.4}
\end{align}
and
$\partial_\mu\xi_\nu+\partial_\nu\xi_\mu+\partial_\mu\partial_\nu\xi$
are the so-called gauge terms (see (\ref{Eqn2.3})) and are
non-covariant depending on the vector $N^\mu$.

The induced Riemann curvature tensor in the leading theory is
given by
\begin{align}\label{Eqn5.5}
\BK{0_-}{R_{\mu\nu\sigma\lambda}(x)}{0_-}^T =
                \BK{0_-}{
                          \partial_\mu \partial_\sigma h_{\nu\lambda}
                          +\partial_{\nu}\partial_{\lambda}h_{\mu\sigma}
                          -\partial_\mu\partial_\lambda h_{\nu\sigma}
                          -\partial_\nu\partial_\sigma h_{\mu\lambda}
                          }
                    {0_-}^T
\end{align}
By substituting the expression (\ref{Eqn5.2}) in (\ref{Eqn5.5}),
we see that  all the terms depending on $\xi^\mu,\xi$
\textit{cancel} in the induced Riemann curvature tensor
$\BK{0_-}{R_{\mu\nu\sigma\lambda}(x)}{0_-}^T$ thus establishing
its covariance. This means that one may restrict
$\BK{0_-}{h_{\mu\nu}(x)}{0_-}^T$ to its covariant
gauge-independent part
\begin{align}\label{Eqn5.6}
\BK{0_-}{h_{\mu\nu}(x)}{0_-}^T &= \left\{ 8\pi G\uI
                                        \int\uD\omega_{\vec{k}}\uE^{\uI kx}
                                                \Big[T_{\mu\nu}(k)
                                                -\frac{\eta_{\mu\nu}}{2}T(k)\Big] +  ~c.c.
                                                \right\}\nonumber\\
                              &\equiv h^o_{\mu\nu}(x)
\end{align}
in applications. The expression for the latter may be further
simplified to
\begin{align}\label{Eqn5.7}
h^o_{\mu\nu}(x) = \left\{ 8\pi G\uI
                            \int (\uD x')
                                \int \uD \omega_{\vec{k}} \uE^{\uI k(x-x')}
                                \Big[T_{\mu\nu}(x') -
                                \frac{\eta_{\mu\nu}}{2}T(x')\Big] +
                                ~c.c.
                    \right\}
\end{align}
The $\vec{k}$-integration as well as the $x^{0'}$-one may be
explicitly carried out leading to
\begin{align}\label{Eqn5.8}
h^o_{\mu\nu}(x) = 2 G
                    \int \frac{\uD^3 \vec{x}'}{|\vec{x} - \vec{x}'|}
                    \Big[T_{\mu\nu} (x^0 - |\vec{x}-\vec{x}'|,\vec{x}')
                    -\frac{\eta_{\mu\nu}}{2}
                    T(x^0 - |\vec{x}-\vec{x}'|,\vec{x}')\Big].
\end{align}

\section{The induced correction to the metric: Application to a Nambu string}\label{Section6}

The metric of spacetime to the leading contribution in our
notation here is defined \cite{Schwinger_1976} by
\begin{align}\label{Eqn6.1}
g_{\mu\nu}(x) = \eta_{\mu\nu} + 2h^o_{\mu\nu}(x),
\end{align}
with the 2 factor, where $h^o_{\mu\nu}(x)$ is given in
(\ref{Eqn5.8}). The leading
contribution to the inverse $g^{\mu\nu}$ is then given by $g^{\mu\nu}=\eta^{\mu\nu}-2h^{o\mu\nu}$.\\
\indent We investigate the contribution to the metric, the induced
geometry and corresponding spacetime measurements due to a string.
The dynamics of the string is described as follows. The trajectory
of the string is described by a vector function
$\vec{R}(\sigma,t)$, where $\sigma$ parametrizes the string. The
equation of motion of the closed sting considered is taken to be
\begin{align}\label{Eqn6.2}
\frac{\partial^2}{\partial t^2}\vec{R}(\sigma,t) -
    \frac{\partial^2}{\partial \sigma^2}\vec{R}(\sigma,t) = 0,
\end{align}
with constraints
\begin{align}\label{Eqn6.3}
\partial_t \vec{R} \cdot \partial_{\sigma}\vec{R} = 0,\quad
    (\partial_t \vec{R})^2
    + (\partial_{\sigma}\vec{R})^2 = 1,\quad
        \vec{R}\Big(\sigma + \frac{2\pi}{\omega},t\Big)
        = \vec{R}(\sigma,t),
\end{align}
for a constant $\omega$. The general solution to (\ref{Eqn6.2}),
(\ref{Eqn6.3}) is given by
\begin{align}\label{Eqn6.4}
\vec{R}(\sigma,t) = \frac{1}{2}
                    \Big[\vec{\Phi}(\sigma - t) +
                        \vec{\Psi}(\sigma + t)\Big],
\end{align}
where $\vec{\Phi},\vec{\Psi}$, in particular, satisfy the
normalization conditions
$(\partial_\sigma\vec{\Phi})^2=(\partial_\sigma\vec{\Psi})^2=1$.
For the system (\ref{Eqn6.2})-(\ref{Eqn6.4}), we consider a
solution of the form
\cite{Manoukian_1991/2,Manoukian_1992,Manoukian_1995,Manoukian_1998}
\begin{align}\label{Eqn6.5}
\vec{R}(\sigma,t) = (\cos \omega\sigma, \sin \omega\sigma,
0)\frac{\sin \omega t}{\omega},
\end{align}
describing a radially oscillating circular string in a plane. The
general expression for the energy-momentum tensor of the string is
given by
\begin{align}\label{Eqn6.6}
T^{\mu\nu}(x) = \frac{M\omega}{2\pi}
                \int_0^{2\pi/\omega} \uD \sigma
        \Big(\partial_t R^{\mu}\partial_t R^{\nu}
                        - \partial_\sigma R^{\mu}\partial_\sigma R^{\nu}\Big)
        \delta^3 \Big(\vec{r} - \vec{R}(\sigma,t)\Big),
\end{align}
where $R^0=t,\ \vec{r}=r(\cos\phi,\sin\phi,0)$, and $M$ provides a
mass scale. The various components of the energy-momentum tensor
are worked out to be
\cite{Manoukian_1991/2,Manoukian_1992,Manoukian_1995,Manoukian_1998}
\begin{align}
T^{00} &= \frac{M}{2\pi r}
            \delta\left(r-\frac{|\sin \omega t|}{\omega}\right)\delta(z),\label{Eqn6.7}\\[0.5\baselineskip]
T^{0i} &= \frac{M}{2\pi r}
            (\cos \phi,\sin \phi, 0)
            \delta\left(r-\frac{|\sin \omega t|}{\omega}\right)\delta(z)
                \cos \omega t\  \textrm{sgn}(\sin \omega t),\label{Eqn6.8}\\[0.5\baselineskip]
T^{11} &= \frac{M}{2\pi r}
            \delta\left(r-\frac{|\sin \omega t|}{\omega}\right)\delta(z)
                [\cos^2 \omega t -\sin^2\phi],\label{Eqn6.9}\\[0.5\baselineskip]
T^{12} &= \frac{M}{2\pi r}
            \delta\left(r-\frac{|\sin \omega t|}{\omega}\right)\delta(z)
                \frac{\sin 2\phi}{2},\label{Eqn6.10}\\[0.5\baselineskip]
T^{22} &= \frac{M}{2\pi r}
            \delta\left(r-\frac{|\sin \omega t|}{\omega}\right)\delta(z)
                [\cos^2 \omega t -\cos^2\phi],\label{Eqn6.11}\\[0.5\baselineskip]
T^{\mu 3} &= 0, \label{Eqn6.12}
\end{align}
where $\textrm{sgn}(\alpha)=\pm1$ for $\alpha\gtrless0$ is the
sign
function, $i=1,2,3$.\\
\indent We note the normalization condition
\begin{align}\label{Eqn6.13}
\int \uD^3 \vec{x}T^{00}(x)= M.
\end{align}
Also for the trace $T^\mu{}_\mu(x)$ of the energy-momentum tensor
we have
\begin{align}\label{Eqn6.14}
T= -\frac{M}{\pi r}\delta\Big(r - \frac{|\sin \omega
t|}{\omega}\Big)\delta(z) \sin^2 \omega t.
\end{align}

It is most interesting to consider spacetime measurements along
the most symmetrical direction in the problem, that is, along the
$z-(x^3-)$ axis perpendicular to the plane of oscillations. Before
doing so, we note that in the plane of oscillations of the string,
$g_{\phi\phi}$ cannot be a function of $\phi$ by symmetry. Also no
cross term $g_{r\phi}$ can occur in this plane, i.e.,
$g_{r\phi}=0$. The metric contributions $h_{rr},h_{00}$, in the
plane of oscillations, are readily obtained. To this end
(\ref{Eqn5.8}), (\ref{Eqn6.7})-(\ref{Eqn6.12}), (\ref{Eqn6.14})
lead for $r\gg1/\omega$
\begin{align}\label{Eqn6.15}
2h_{11}(x) &\simeq \frac{4G}{r}
                \int \uD^3 \vec{x}'
                \Big[T_{11}(x^0-r,\vec{x}') - \frac{T(x^0
                -r,\vec{x}')}{2}\Big]
                = \frac{2GM}{r}\nonumber\\
            &\simeq 2 h_{22}(x), \qquad h_{12} \simeq 0
\end{align}
where $1/\omega$ is the maximum radial extension of the string.
Using the identity $h_{rr}=\cos^2\phi\, h_{11}+\sin^2\phi\,
h_{22}+\sin 2\phi \,h_{12}$, it leads to
\begin{align}\label{Eqn6.16}
g_{rr} \simeq \Big(1 + \frac{2GM}{r}\Big).
\end{align}
On the other hand,
\begin{align}\label{Eqn6.17}
2h_{00}(x)  &\simeq \frac{4GM}{r}
                \int\uD^3 \vec{x}'
                \Big[T_{00}(x^0 -r, \vec{x}') +
                \frac{T(x^0-r,\vec{x}')}{2}\Big]\nonumber\\
            &=\frac{4GM}{r}\cos^2 \omega (t-r)
\end{align}
or
\begin{align}\label{Eqn6.18}
g_{00}(x) \simeq -\Big(1 - \frac{4GM}{r}\cos^2 \omega (t-r)\Big)
\end{align}
where we recall that the Minkowski metric is taken to be
$[\eta_{\mu\nu}]$=diag[-1,1,1,1].

For an observer at a fixed $r\gg1/\omega$ in the plane of
oscillations of the string, then time slows down by a factor
\begin{align}\label{Eqn6.19}
\frac{1}{(T_2 - T_1)}
            \int_{T_1}^{T_2} \sqrt{-g_{00}}\uD t =
            1- \frac{GM}{r}\left\{1 + \cos \omega(T_1 + T_2 -2r)
            \frac{\sin \omega(T_2 -T_1)}{\omega(T_2 -
            T_1)}\right\},
\end{align}
relative to a time lapsed of length $(T_2-T_1)$ in empty space.

For spacetime measurements along the z-axis, we have explicitly
\begin{align}\label{Eqn6.20}
2h^o_{33}(x) = 4GM
                \int_0^{\infty}
                \frac{\uD r'}{\sqrt{r'^2 + z^2}}
                \delta\left(r' - \frac{\sin \omega(t - \sqrt{r'^2 +
                z^2})}{\omega}\right)r'^2\omega^2
\end{align}
Again, since $r'$ does not exceed $1/\omega$, we have  for an
observer at $|z| \gg 1/\omega$
\begin{align}\label{Eqn6.21}
g_{33}(x) \simeq 1 + \frac{4GM}{|z|}\sin^2 \omega (t- |z|),
\end{align}
showing an interesting oscillatory behaviour in the space metric
with a relative expansion of length.\\

Similarly, we obtain
\begin{align}\label{Eqn6.22}
g_{00}(x) \simeq -\left(1- \frac{4GM}{|z|}\cos^2
\omega(t-|z|)\right).
\end{align}

\section{Conclusion}\label{Section7}
The positivity constraint as well as the spin content of the
theory of gravitons interacting with \textit{a priori}
non-conserved external energy-momentum tensor was established. As
emphasized throughout, relaxing this conservation law is necessary
so that variations of the ten components of the energy-momentum
tensor may be varied independently which goes to the heart of the
functional differential formalism of quantum field theory. The
expectation value formalism of the theory within the above context
was derived at finite temperature for gravitons. Thermal averages
of the generated gravitational field and their correlations may be
then obtained by functional differentiations of the resulting
generating functional at finite temperature which coincide with
the corresponding expectation values $\bra{0_-}\cdot\ket{0_-}$ at
zero temperature. The covariance of the induced Riemann curvature
tensor was established in spite of the gauge constraint which
ensures only two polarization states of the graviton. An
application was carried out to determine the induced correction to
the Minkowski metric resulting from a closed string arising from
the Nambu action as a solution of a circularly oscillating string.
Radiative corrections play an important role as the induced
geometry may, in general, depend on temperature. Technically, this
may be seen as follows. The multiplicative factor in the
generating functional $\mathscr{F}[T^1,T^2;\tau]$ in
(\ref{Eqn4.16}) depending on temperature is given by
\begin{align}\label{Eqn7.1}
\exp \left[-8\pi G
                \int (\uD x)(\uD x')
                \Big(T^1_{\mu\nu}(x) -
                    T^2_{\mu\nu}(x)\Big)
                D^{\mu\nu,\sigma\rho}(x,x';\tau)
                \Big(T^1_{\sigma\rho}(x')- T^2_{\sigma\rho}(x')\Big)
                \right]
\end{align}
where $D^{\mu\nu,\sigma\rho}(x,x';\tau)$ is defined in
(\ref{Eqn4.17}), (\ref{Eqn3.28})-(\ref{Eqn3.30}). Consider a
familiar correction to the leading order in the Lagrangian density
given by
$h^{\mu\nu}(x)\left(\tau_{\mu\nu}\right.$\\$+T^{(m)}_{\mu\nu})$,
where $\tau_{\mu\nu}$, $T^{(m)}_{\mu\nu}$ are energy-momentum
tensors of the gravitational field and matter, respectively. For
example, if $T^{(m)}_{\mu\nu}$ corresponds to a real scalar field
coupled in turn to an external source $K(x)$, then the
multiplicative factor in the corresponding generating functional
of the scalar field depending on temperature is clearly given by
\begin{align}\label{Eqn7.2}
\exp\left[- \int(\uD x)(\uD x')
            \Big(K^1(x) - K^2(x)\Big)
                \Delta^+(x,x';\tau)
            \Big(K^1(x') - K^2(x')\Big)\right]
\end{align}
where
\begin{align}\label{Eqn7.3}
\Delta^+ (x,x';\tau) =
                    \int\frac{\uD^3 \vec{k}\uE^{\uI k(x-x')}}{(2\pi)^3 2\sqrt{\vec{k}^2 + m^2}}
                    \Big(\uE^{\beta\sqrt{\vec{k}^2 + m^2}} - 1\Big)^{-1},
\end{align}
$k^0=+\sqrt{\vec{k}^2+m^2}$, and $m$ is the mass of the scalar
field. Now both $\tau_{\mu\nu}$ and $T^{(m)}_{\mu\nu}$ are
\textit{quadratic} in their respective fields. To generate the
term $h^{\mu\nu}\tau_{\mu\nu}$, we then need to functionally
differentiate (\ref{Eqn7.1}), say, with the external source
$T^1_{\mu\nu}$ \textit{three} times, also additively w.r.t.
$T^2_{\mu\nu}$ according to the quantum dynamical principle
\cite{Schwinger_1961,Manoukian_1987,Manoukian_1991/1}. On the
other, hand to generate $T^{(m)}_{\mu\nu}$, we have to
functionally differentiate (\ref{Eqn7.2}) twice with respect to
the external sources $K^{1,2}$ of the scalar field. Finally to
generate the thermal average of $h_{\mu\nu}$, we have to
functionally differentiate once more w.r.t. $T^1_{\mu\nu}$ and
then set $T^1_{\mu\nu}=T^2_{\mu\nu}\equiv T_{\mu\nu}$, and
$K^1=K^2\equiv K$. That is, all in all, we have an \textit{even}
number of functional differentiations w.r.t. the corresponding
external sources to generate the thermal average
$\avg{h_{\mu\nu}}^T_\tau$ before setting the equality of the
sources just mentioned and thus generate a temperature dependence
in $\avg{h_{\mu\nu}}^T_\tau$. This is unlike the situation in the
leading order in which we have to differentiate only once w.r.t.
$T^1_{\mu\nu}$ to generate $\avg{h_{\mu\nu}}^T_\tau$ before
setting $T^1_{\mu\nu}-T^2_{\mu\nu}=0$, resulting no temperature
dependence in the former expression as seen in (\ref{Eqn5.1}). The
study of higher orders, however, requires a detailed analysis of
Faddeev-Popov-like factors of the type discovered in
\cite{Limboonsong_2006}, \cite{Manoukian_2007},
 as generated in the functional differential treatment (see Sect.3 in \cite{Manoukian_2007},
\cite{Manoukian_1986/1}, \cite{Limboonsong_2006},
\cite{Sukkhasena_2007}) which would in turn lead to extra vertices
coming from the second term on the right-hand side of
(\ref{Eqn1.3}) and its generalizations and complicates matter
quite a bit in gravitation. This formidable problem as well as
convergence aspects \cite{Manoukian_1983} will be investigated in
a future report. Physically, temperature dependence of the
underlying induced geometry is also clear. When we perform a
thermal average, we introduce in the process, a
\textit{background} of gravitons, and in general other particles
depending on the matter fields considered. These particles in turn
would then act as additional \textit{sources} of gravitation
contributing to the net induced gravitational field  and this
happens \textit{only} when non-linearities as field
\textit{interactions} are considered, and corresponding radiative
corrections are taken into account.

\end{document}